\documentclass[apj]{emulateapj}

\newcommand  \kms      {\ifmmode {\rm km\,s}^{-1} \else km\,s$^{-1}$\fi}

\newcommand  \cmii     {\hbox{cm$^{-2}$}}
\newcommand  \ergs     {\ifmmode {\rm ergs\,s}^{-1} \else ergs s$^{-1}$\fi}
\newcommand  \ergcms   {\ifmmode {\rm ergs\,cm}^{-2}\,{\rm s}^{-1}
                        \else ergs\,cm$^{-2}$\,s$^{-1}$\fi}
\newcommand  \ergcmsA {\ifmmode{\rm ergs\,cm}^{-2}\,{\rm s}^{-1}\,{\rm\AA}^{-1}
                        \else ergs\,cm$^{-2}$\,s$^{-1}$\,\AA$^{-1}$\fi}
\newcommand \ergcmsHz {\ifmmode{\rm ergs\,cm}^{-2}\,{\rm s}^{-1}\,{\rm Hz}^{-1}
                        \else ergs\,cm$^{-2}$\,s$^{-1}$\,Hz$^{-1}$\fi}
\newcommand  \phcms    {\ifmmode {\rm ph\,cm}^{-2}\,{\rm s}^{-1}
                        \else ,ph\,cm$^{-2}$\,s$^{-1}$\fi}
\newcommand  \phcmsA   {\ifmmode {\rm ph\,cm}^{-2}\,{\rm s}^{-1}\,{\rm\AA}^{-1}
                        \else ph\,cm$^{-2}$\,s$^{-1}$\,\AA$^{-1}$\fi}
\newcommand{\mbh}{$M_{\rm BH}$}

\newcommand{\msig}{$M-{\sigma*}$}
\def\micron{\ifmmode \mu{\rm m} \else $\mu$m\fi}
\def\kms{\ifmmode {\rm km\,s}^{-1} \else km\,s$^{-1}$\fi}
\def\Hubble{\ifmmode {\rm km\,s}^{-1}\,{\rm Mpc}^{-1}
        \else km\,s$^{-1}$\,Mpc$^{-1}$\fi}
\def\ergsec{\ifmmode {\rm ergs\;s}^{-1} \else ergs s$^{-1}$\fi}
\def\ergscm{\ifmmode {\rm ergs\,s}^{-1}\,{\rm cm}^{-2}
          \else ergs\,s$^{-1}$\,cm$^{-2}$\fi}
\def\ergscmA{\ifmmode {\rm ergs\,s}^{-1}\,{\rm cm}^{-2}\,{\rm \AA}^{-1}
          \else ergs\,s$^{-1}$\,cm$^{-2}$\,\AA$^{-1}$\fi}
\def\ergscmHz{\ifmmode {\rm ergs\,s}^{-1}\,{\rm cm}^{-2}\,{\rm Hz}^{-1}
          \else ergs\,s$^{-1}$\,cm$^{-2}$\,Hz$^{-1}$\fi}
\def\rblr{\ifmmode r_{BLR} \else $r_{BLR}$\fi}
\def\Msun{\ifmmode M_{\odot} \else $M_{\odot}$\fi}
\def\msun{\ifmmode M_{\odot} \else $M_{\odot}$\fi}
\def\Lsun{\ifmmode L_{\odot} \else $L_{\odot}$\fi}
\def\qo{\ifmmode q_{0} \else $q_{0}$\fi}
\def\Ho{\ifmmode H_{0} \else $H_{0}$\fi}
\def\ho{\ifmmode h_{0} \else $h_{0}$\fi}
\def\qo{\ifmmode q_{0} \else $q_{0}$\fi}
\def\ao{\ifmmode a_{0} \else $a_{0}$\fi}
\def\to{\ifmmode t_{0} \else $t_{0}$\fi}
\def\Halpha{\ifmmode {\rm H}\alpha \else H$\alpha$\fi}
\def\Hbeta{\ifmmode {\rm H}\beta \else H$\beta$\fi}
\def\hb{\ifmmode {\rm H}\beta \else H$\beta$\fi}
\def\Hgamma{\ifmmode {\rm H}\gamma \else H$\gamma$\fi}
\def\Hdelta{\ifmmode {\rm H}\delta \else H$\delta$\fi}
\def\Lya{\ifmmode {\rm Ly}\alpha \else Ly$\alpha$\fi}
\def\Lyb{\ifmmode {\rm Ly}\beta \else Ly$\beta$\fi}
\def\hi{\ifmmode \mbox{{\rm H}\,{\sc i}} \else H\,{\sc i}\fi}

\def\ciii{\ifmmode {\rm C}\,{\sc iii} \else C\,{\sc iii}\fi}
\def\civ{C\,{\sc iv}\,$\lambda1549$}

\def\lnii_ha{L([N\,{\sc ii}])/L(H$_{\alpha}$) }
\def\loiii_hb{L([O\,{\sc iii}])/L(H$_{\beta}$) }

\def\o5007{[O\,{\sc iii}]\,$\lambda5007$}

\def\ne212m {[Ne\,{\sc ii}]\,$12.8 \mu m$}

\def \Lop{$L_{5100}$}
\def \L5100{$L_{5100}$}
\def \Lbol{$L_{bol}$}
\def \lbol{$L_{bol}$}
\def \Ledd{$L/L_{\rm Edd}$}

\def  \Ncol        {\hbox{$ {N_{\rm col}} $}}      

\def  \kms         {\hbox{km s$^{-1}$}}          
\def  \ergs        {\hbox{ergs s$^{-1}$}}              

\def  \cmii        {\hbox{cm$^{-2}$}}

\def  \La          {\ifmmode {\rm Ly}\alpha \else Ly$\alpha$\fi}
\def  \Ka          {\ifmmode {\rm K}\alpha \else K$\alpha$\fi}
\def  \Lb          {\ifmmode {\rm L}\beta \else L$\beta$\fi}
\def  \Ha          {\ifmmode {\rm H}\alpha \else H$\alpha$\fi}
\def  \Hb          {\ifmmode {\rm H}\beta \else H$\beta$\fi}
\def  \hb          {\ifmmode {\rm H}\beta \else H$\beta$\fi}
\def  \Pa          {\ifmmode {\rm P}\alpha \else P$\alpha$\fi}
\def  \CIIIb       {\ifmmode {\rm C}\,{\sc iii]}\,\lambda1909
                     \else C\,{\sc iii]}\,$\lambda1909$\fi}
\def  \CIV         {\ifmmode {\rm C}\,{\sc iv}\,\lambda1549
                     \else C\,{\sc iv}\,$\lambda1549$\fi}
\def  \MgII         {\ifmmode {\rm Mg}\,{\sc ii}\,\lambda2798
                     \else Mg\,{\sc ii}\,$\lambda2798$\fi}
\def  \OVI         {\ifmmode {\rm O}\,{\sc vi}\,\lambda1035
x
                     \else O\,{\sc vi}\,$\lambda1035$\fi}

\shorttitle{Radiation pressure in AGN}
\shortauthors{Netzer, Marziani}

\begin{document}

\title{The effect of radiation pressure on emission line profiles and black hole mass determination
 in active galactic nuclei}

\author{
Hagai Netzer\altaffilmark{1},
Paola Marziani\altaffilmark{2}
}

\altaffiltext{1} {School of Physics and Astronomy and the Wise
  Observatory, The Raymond and Beverly Sackler Faculty of Exact
  Sciences, Tel-Aviv University, Tel-Aviv 69978, Israel}
\altaffiltext{2}{INAF, Osservatorio Astronomico di Padova, Vicolo dell' Osservatorio 5, IT35122 Padova, Italy} 

\begin{abstract}

We present a new analysis of the motion of pressure-confined, broad line region (BLR) clouds in active galactic nuclei (AGNs)
taking into account the combined influence of gravity and radiation pressure. We calculate 
cloud orbits under a large range of conditions and include the effect of column density variation as a function
of location. The dependence of radiation pressure force on the level of  ionization and the column density are
accurately computed.
The main results are:
a. The mean cloud locations (\rblr) and line widths (FWHMs) are combined in such a way that the simple virial mass estimate,
$r_{BLR} FWHM^2/G$, gives a reasonable approximation to \mbh\ even when radiation pressure force is important.
The reason is that $L/M$ rather than $L$ is the main parameter affecting
the planar cloud motion.
b. Reproducing the mean observed \rblr, FWHM and line intensity of \hb\ and \civ\ requires at least  
two different populations of clouds.
c. The cloud location is a function of both $L^{1/2}$ and $L/M$. Given this, we suggest a new approximation for
\rblr\ which, when inserted into the BH mass equation, results in a new approximation for \mbh.
The new expression involves
$L^{1/2}$, FWHM and two constants that are obtained from a comparison with available
\msig\ mass estimates.
It deviates only slightly from the old mass estimate at all luminosities.
d. The quality of the present black hole mass estimators depends, critically, on the way the present \msig\ AGN sample (29 objects)
represents the overall population, in particular the distribution of \Ledd.

\end{abstract}


\section{Introduction}
The profiles of the broad emission lines in the spectrum of active galactic nuclei (AGNs)
 are the main source of information about the motion of the high density gas in the broad line region (BLR).
 Detailed studies of such profiles  have been the focus of intense investigation for many years 
 (see Netzer 1990 for a review of older work and Marziani et al. 1996 and Richards et al. 2002
 for more recent publications). 
Unfortunately, several rather different geometries can 
 conspire to result in similar line profiles and today, there is no way to infer, directly, the global 
 BLR motion from line profile fitting.

 A less ambitious  goal is to use a measure of the observed line width, e.g. the line
FWHM, or the line dispersion (see Peterson et al. 2004 for definitions) as indicators of the mean emissivity-weighted velocity of the BLR gas. Such measurements
 are crucial for deducing black hole (BH) mass (\mbh) in cases where the emissivity-weighted radius,
 \rblr, is measured directly from reverberation mapping (RM) experiments, or 
estimated from from  L-$r_{BLR}$  relationships that are based on such studies 
 (see Kaspi et al. 2000; Kaspi et al 2005; Vestergaard and Peterson 2006 for reviews). 
 A typical expression of this type  is
\begin{equation} 
r_{BLR}=a \left[ \frac { L_{5100} }{ 10^{46}\, {\rm erg\,s^{-1} } } \right] ^{\gamma} \,\, pc,
\label{eq:old_rblr}
\end{equation}
where \Lop\ is the continuum luminosity ($\lambda L_{\lambda}$) at 5100\AA\
and $\gamma = 0.6 \pm 0.1$. The constant $a$ depends on the line in question. For \hb,
$a \simeq 0.4$\,pc (e,g, Bentz et al. 2009) and for \civ, 
$a \simeq 0.13$\,pc (Kaspi et al. 2007 after assuming $L_{1350}=2 L_{5100}$). For a virialized BLR,
the above \rblr\ can be combined with a measure of the FWHM, or the line dispersion, to obtain
the BH mass, 
\begin{equation}
 M_{BH}=f r_{BLR} FWHM^2/G \,
\label{eq:old_mbh}
\end{equation}
where the constant $f$ is a geometrical correction factor of order unity that takes into accounts 
the (unknown) gas distribution and dynamics. 
Various possible values of $f$ have been computed by Collin et al. (2006) for various possible geometries.
However, the only empirical way to determine $f$ is to compare the results of
eqn.~\ref{eq:old_mbh} with independent measurements of \mbh, like those available in the the case where 
the central BH resides in a bulge and \mbh\ can be estimates from the \msig\ relationship (e.g. Tremaine et al. 2002).
Such comparisons by Onken et al (2004) and by Woo et al. (2010), using the \hb\ RM data base, suggest $f=1 \pm 0.1$.

In a  recent paper,  Marconi et al. (2008; hereafter M08)
 investigated  the role of radiation pressure force and its effect on the motion of the BLR
gas and the required modification to the BH mass estimate.
According to M08, radiation pressure plays an important
role in affecting the cloud motion provided the column density (\Ncol) of most BLR clouds is  smaller
than about $10^{23}$ \cmii. According to M08, in such a case, there is a need to add a second term
to eq.~\ref{eq:old_mbh}. This term  depends on the source 
luminosity and \Ncol.
 The modified form suggested in M08 is
\begin{equation}
 M_{BH}=f_1 r_{BLR} FWHM^2/G + f_g L/N_{col}\,
\label{eq:M08_mbh}
\end{equation}
where$f_1$ replaces $f$ in eqn.~\ref{eq:old_mbh} 
 and $f_g$ is a second constant. If $L=L_{5100}/10^{44} \, {\rm erg\,s^{-1}}$ and \Ncol\ is measured in units of $10^{23}\, {\rm cm^{-2}}$,
$f_g \simeq 10^{7.7}$\Msun\..
 According to M08, failing to account for the second term results in the underestimation of \mbh. 
Obviously, the inclusion of such a term results in $f_1<f$.
M08 repeated the analysis of Onken et al. (2004) and Vestergaard and Peterson (2006), taking into account
the new term and solving for $f_1$ and \Ncol. This resulted in  $f_1 \simeq 0.56$ and \Ncol$\simeq 10^{23}$
\cmii. 

The M08 suggestion can be tested by comparing low redshift samples of type-I and type-II AGNs
since the estimate of \mbh\ in the latter does not involve the source luminosity and gas dynamics.
Netzer (2009) carried out such a comparison and found that radiation pressure force plays only a marginal role in such sources. The conclusion is that, in many AGNs, the mean column density 
of the BLR clouds exceeds $\sim 10^{23}$\cmii. In a later work, Marconi et al (2009; hereafter M09)
 argued that firm conclusions
regarding the role of radiation pressure force are difficult to obtain since the column density in some BLRs
can be different than in others and there is no simple way to evaluate the overall effect of such a column
density distribution. The treatment of a certain type of cloud in all sources, or
even in a single BLR, is of course highly simplified and 
 eqns.~\ref{eq:old_mbh} and ~\ref{eq:M08_mbh} must be  treated as crude first approximations.

The critical and detailed evaluation of the role of radiation pressure force in  ``real'' BLRs is the subject of the present paper. In \S2 we
present our basic equations and in \S3 we use them to calculate various expected broad emission line 
profiles and 
mass normalization factors, $f$.  \S4 deals with the evaluation of present day 
\mbh\ estimates and suggests a new way to estimate \mbh\ and \rblr\ 
which is consistent with our calculations.

\section{Cloud motion in the BLR}
In this work we focus on the ``cloud model'' of the BLR. The general framework of this model is explained
in Netzer (1990) and in Kaspi and Netzer (1999) and a major empirical justification is obtained from 
the recent X-ray detected single blobs, or clouds, moving in a region which is typical, in terms
of velocity and dimension, of the BLR (Risaliti et al. 2010; Maiolino et al. (2010).
We do not consider the locally optimally-emitting cloud (LOC) 
model (Baldwin et al. 1995; Korista \& Goad 2000) where, at every location, there is a large range in cloud properties.
The dynamics of the BLR gas in this model has never been treated and is far more complicated than the one considered here.
Another possibility that has been discussed, extensively, is that wind from the
inner disk plays an important role in feeding and driving the
BLR gas. Possible evidence for this scenario comes from radio observations (e.g. 
Vestergaard, Wilkes, \& Barthel 2000; Jarvis \& McLure 2006)  and
spectropolarimetry (Smith et al. 2004; Young et al. 2007). Theoretical considerations
are discussed in 
Bottorff et al. (1997), Murray \& Chiang (1997),  Proga, Stone, \& Kallman (2000),
 Everett (2003),  Young et al. (2007), and several other papers. 
While our calculations apply to any cloud, even those created and driven by such winds,
the specific examples given below are more applicable to bound clouds where inward and outward motions are
both allowed.

\subsection{The equation of motion of BLR clouds}

The basic equation of motion, ignoring drag force, is 

\begin{equation}
a(r) = \frac{\sigma_T L_{bol}}{\mu m_H c 4 \pi r^2}
 [M(r)-1/\Gamma] -\frac{1}{\rho} \frac{dP_g}{dr} \,\, ,
\label{eq:a_r_modified_1}
\end{equation}
where $M(r)$ is the force multiplier, $L_{bol}$ is the bolometric luminosity,
$\mu$ is the average number of nucleons per electron,
 and $\Gamma=$\Ledd. The force multiplier depends on
 the gas composition and its level of ionization.  
 An interesting case is a Compton thin neutral cloud that absorbs all the ionizing radiation (a 
Compton thin ``block''). In this case 
$M(r) \simeq \alpha(r) /(\sigma_T  N_{col}$) where $N_{col}$ is the hydrogen column density and
 $\alpha(r)$ is the fraction of the bolometric luminosity which is absorbed by the
gas. For such a ``block'', $\alpha(r)=L_{ion}/L_{bol}$ but in general $\alpha(r)$ 
 is radius dependent because of the changing column density and level of ionization of
the gas (see below).

Ignoring thermal pressure we obtain
\begin{equation}
 a(r)=\frac{L_{bol} }{r^2} \left[ \frac{ 1.14 \times 10^{-11} \alpha(r) }{ N_{23} }
    - \frac{ 8.8 \times 10^{-13} }{ \Gamma } \right ]
\label{eq:a_r_modified_2}
\end{equation}
where $N_{23}=N_{col}/10^{23}$.
Thus, radiation pressure is the dominant force when
\begin{equation}
 \Gamma \geq 7.7 \times 10^{-2} \frac { N_{23} }{ \alpha(r) } \,\, .
\label{eq:lim_Gamma}
\end{equation}
The above expressions, including the one for the limiting $\Gamma$, include only radial terms and assume a pure radially dependent radiation pressure
force. The calculations of real orbits, and the conditions for cloud escape, require their integration and will thus include the standard 
constants of motion (energy and angular momentum). Obviously, the conditions for escape depend on the cloud azimuthal velocity, $v_{\theta}$,
and can differ substantially from what is obtained by using eqn.~\ref{eq:lim_Gamma}. 

M08 and M09  derived similar expressions for the case of completely opaque clouds. 
According to them, radiation pressure dominates the cloud motion if
\begin{equation}
 \Gamma \ge 1.27 \times 10^{-2} b_{5100} N_{23}
\end{equation}
where $b_{5100}=L_{bol}/L_{5100}$. The two expressions 
provide the same limiting $\Gamma$ when 
\begin{equation}
 \alpha(r) b_{5100} \simeq 6.1 \,\, .
\end{equation}
A recent paper by  Ferland et al. (2009), where  mostly neutral, infalling clouds are considered,
reaches basically identical conclusions.

\subsection{Confined clouds}

BLR clouds are likely to be confined.
The confining mechanism is not known but high temperature gas and magnetic confinement have been proposed.
The approach chosen here is consistent with the idea of magnetic confinement and some justifications for it are
given in Rees, Netzer and Ferland (1989). We adopt a simple model of numerous individual clouds that are moving under
the combined influence of the BH gravity and radiation pressure force. Following Netzer (1990), and Kaspi and Netzer (1999),
we assume that clouds retain their mass as they move in or out and the gas density changes with radius in
a way that depends on the radial changes of the confining pressure.

Assume the external pressure and hence the gas density in individual clouds are proportional to the radial coordinate,
$n_H \propto r^{-s}$. 
A reasonable guess that agrees with observations is $ 1 \leq s \leq 5/2$ (Rees, Netzer and Ferland 1989).
This results in a radial dependence of the ionization parameter
(the ratio of ionizing photon density to gas density), $U \propto r^{s-2}$. For spherical clouds,
$N_{col} \propto r^{-2s/3}$,
$R_c \propto r^{s/3}$ and 
$A_c \propto r^{2s/3}$, 
where  $R_c$ is the cloud radius and $A_c$ its geometrical cross section. 
The line intensity contributed by a single cloud, $\epsilon(r)$, depends on its covering factor and the line emissivity
$j(r)$ which depends on the conditions in the gas,
\begin{equation}
   \epsilon(r) \propto j(r) A_c/r^2 \propto j(r) r^{2s/3-2} \,\,.
\label{eq:emissivity}
\end{equation}
In the real calculations we ignore factors of order unity relating the mean cloud ``size'' and its mean column density
since this is not known and require different type of calculations.

The above considerations suggest that
the importance of radiation pressure increases with distance because of the dependence of $N_{col}$ 
on $r$, i.e. 
\begin{equation}
 \Gamma_{lim} \propto r^{-2s/3}/\alpha(r) \,\, .
\end{equation}
Since  $\Gamma$ depends on the global accretion rate which has little to do with cloud properties, 
the more physical approach is to consider the case of a certain  $\Gamma$ and follow the cloud motion.
The examples discussed below follow this approach.

In this work we consider three types of clouds:
\begin{enumerate}
 \item 
Very large column density clouds where radiation pressure force is negligible at all distances.
Here virial cloud motion is a good approximation (the Ferland et al 2009 infalling clouds belong to
this category). 
 \item 
Cloud for which radiation pressure is very important somewhere inside the ``classical BLR'' (e.g. inside
 the RM radius). Such clouds
 will escape the system on dynamical time scales and their
  contribution to the line profiles is small except for times immediately after a large increase in \Lbol.
 \item
 Clouds for which radiation pressure is non-negligible but is not strong enough to allow escape.
 Such clouds are the ones discussed by M08  albeit without the radial dependence of \Ncol\ considered here.
  This case is the one most relevant to real BLRs and we discuss it
  in detail in the following section.
\end{enumerate}

\subsection{Modified equation of motion}

The modified equation of motion is obtained from eqn.~\ref{eq:a_r_modified_2} by including the
radial dependence of $N_{col}$. We define 
 $r_{23}$ to be  the distance where $N_{23}=1$. This gives 
\begin{equation}
 a(r)=\frac{L_{bol}}{r^2} \left[ \frac{ 1.14 \times 10^{-11} \alpha(r) }{ (r/r_{23})^{-2s/3} }
    - \frac{ 8.8 \times 10^{-13} }{ \Gamma } \right ] \,\, .
 \label{eq:a_r_modified_3}
\end{equation}
The column density dependent  critical distance where radiation force
dominates the cloud motion  is,
\begin{equation}
 \frac {r}{r_{23}} \geq \left[ \frac{7.7 \times 10^{-2} }{ \alpha(r) \Gamma} \right] ^{3/2s} \,\, .
 \label{eq:r_min}
\end{equation}
For example, the case of $s=1$ and $\alpha(r)=0.5$ gives
a critical radius of $r=0.06 \Gamma^{-1.5} r_{23}$ for radially moving clouds.
This dependence of $r$ on $\Gamma$ is the main motivation to suggest a new method for evaluating \mbh\
and \rblr\ (\S4).  As explained, the critical radius should not be confused with the point of escape from the system.
Non-radial velocity components ($v_{\theta}$), that reflect
the energy and angular momentum of the system will act to reduce this radius (see examples below).

The motion of BLR clouds  with the above properties involves an 
acceleration term of the form,
\begin{equation}
 a(r)=\frac {c_1 \alpha(r) }{ r^{2-2s/3} } - \frac{ c_2}{r^2 } \,\, ,
\label{eq:a_r_again}
\end{equation}
where
\begin{equation}
 c_1=1.14 \times 10^{-11} L_{bol} r_{23}^{-2s/3}
\end{equation}
and
\begin{equation}
 c_2=8.8 \times 10^{-13}L_{bol}/\Gamma \,\, .
\end{equation}
The radial  potential is,
\begin{equation}
 \Phi(r)= - \int_r^{r*} a(r) dr  \,\, ,
\label{eq:potential}
\end{equation}
where $r*$ is the radius where $\Phi(r)=0$.
Below we use this potential to calculate cloud
 orbits and  line profiles. The energy and angular momentum terms that result from the above integration, are included in the calculation by fixing
the initial conditions, $r$ and the two velocity components at this location. 

\section{Line profile calculations}
\subsection{Method}
We carried out a series of calculations 
under a variety of conditions considered to be typical of different BLRs. Every model is calculated for 
 assumed \mbh\ and  $\Gamma$. This specify \Lbol\ and thus the potential $\Phi(r)$.
The additional model parameters are:
\begin{enumerate}
 \item 
 The radial parameter $s$. 
 \item
 The cloud column density normalization factor $r_{23}$. 
 \item
 The initial radius $r_0$ and the initial velocity $v_0=v_{\theta}(r_0)$.
 We assume that the orbits of clouds with very large $N_{col}$ are ellipses of given
 eccentricities. $r_0$ is chosen to be the apogee of the orbit and $v_0$ (given below in
 units of the Keplerian velocity, $v_{Kepler}$) is determined from these conditions. This a simple
way to specify the angular momentum.
In the examples below, we focus  on those cases where the resulting FWHMs are consistent with the observations
 of the broad \hb\ and \civ\ emission lines but give details for several others.
  \item 
 The initial ionization parameter, $U(r_0)$. We note that the exact value of the gas density, $n_H(r)$,
 is less important. In the following we assume 
  $n_H(r_0)=10^{10}$ \cmii\ for all cases.
 \item 
 The three-dimensional distribution of orbits. This is done in two steps. First
 we calculate the motion of numerous identical clouds in a plane and then distribute many
 such planes in a spherical geometry specified by the inclinations of the planes
to the line of sight. The profiles given below are only those for a line of sight which is perpendicular to the central
plane of motion (if such a plane exists). All calculations assume a large enough number of clouds such that the
predicted profiles are smooth (see Bottorff and Ferland 2000 for discussion and earlier references on this issue).
\end{enumerate}
Physical properties that {\it are not} included in the present calculations are non-isotropic central radiation field, 
non-isotropic line emission,
the photoionization of gas with a range of density and metallicity, 
different inclinations of the line of sight to
the central plane of motion, large cloud covering factors in a specific direction, 
and central obscuration, e.g. by the accretion disk.
Several of those are likely to be important in real BLRs but are beyond the scope of the present work.

\begin{figure}
 \plotone{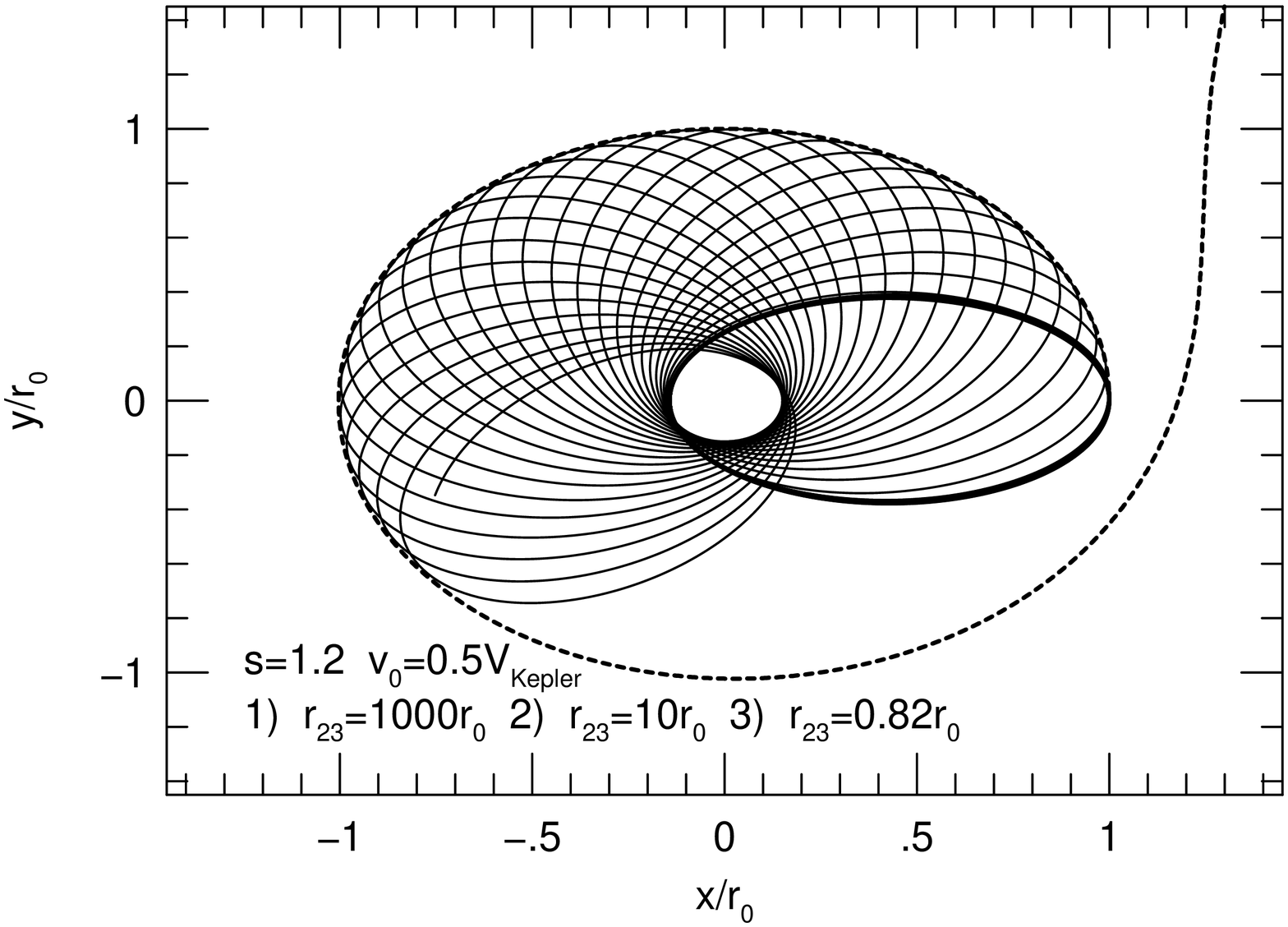}
 \caption{Planar orbits of three clouds with $\Gamma=0.1$ and different column densities.
  The large column
 density cloud (thick line, $r_{23}=1000 r_0$) moves in a closed elliptical orbit. A smaller column density
 cloud (thin line, $r_{23}=10 r_0$) moves in a closed rotating orbit and a marginal column density cloud
(dashed line, $r_{23}=0.82r_0$) escapes the system.
 }
 \label{fig:motion_plane}
\end{figure}

Fig.~\ref{fig:motion_plane} illustrates the orbits of three $s=1.2$, $r_0=10^{17}$ cm,
$\Gamma=0.1$ and $v_0=0.5 v_{Kepler}$ clouds moving under the influence of a $10^8$ \Msun\ BH.
The first is an ellipse typical of a cloud which is  not affected
by radiation pressure (e.g. $r_{23}=1000 r_0$). This is shown by a thick solid line. Formally speaking,
such clouds are Compton thick but this is of no practical implications since the only intention is to
show a simple, gravity dominated orbit.
The second is a case where  $r_{23}=10 r_0$. Here, radiation pressure force
is significant and acts to constantly changing the direction of motion
of the cloud. This results
in a rotating planar orbit. The third orbit (dashed line) follows the trajectory of a smaller column 
density cloud ($r_{23}=0.82 r_0$) that escapes the system.
Increasing $\Gamma$ will result in similar type orbits for the rotating orbit second cloud except that the angle between
two successive revolutions will increase.

\begin{figure}
 \plotone{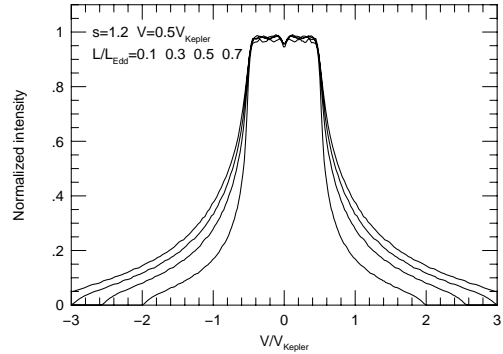}
\caption{Line profiles for  spherical s=1.2 atmospheres around a $10^8$\msun\ BH and a
range of $\Gamma$ as marked. All clouds start at $r_0=10^{17}$ cm with $v_r=0$ and $v_{\theta}=0.5\,v_{Kepler}$.
The column densities are changing as $(r/r_{23})^{-2s/3}$ with
$r_{23}=10^{18}$ cm ($N_{col} \approx 6.3 \cdot 10^{23}$ cm$^{-2}$  at $r_{0}$). The FWHM of the profile decreases with the increasing $\Gamma$ due to the increasing importance of radiation pressure force. The profile parameters are listed in Table~1.
}
\label{fig:profiles_Ledd}
\end{figure}

We calculated various line profiles for the case of \mbh=$10^8$\Msun, $r_{23}=10 r_0$, $v_0=0.5$
 and $\Gamma$ in the range of 0.05 (negligible radiation pressure force) to
 0.735 (just below escape). The bolometric luminosity in each of those is obtained from the combination of
\mbh\ and $\Gamma$.
We assume $\alpha(r)=0.5$ at all radii and $\epsilon(r)$ which takes into
 account {\it only} geometrical factors (i.e. constant $j(r)$, see eqn.~\ref{eq:emissivity}) and isotropic line emission.
  In terms of total line emission, this  is a reasonable approximation for lines like \hb\ that reprocess roughly a constant fraction of the ionizing
 continuum radiation. Obviously, a large optical depth in \hb\ will result in line emission anisotropy which is not considered here.
It is not appropriate for lines like \civ\ whose intensity is more sensitive to
 the level of ionization and the gas temperature. At this stage we specifically avoid the use of a varying  $\alpha(r)$
since the effect on the orbit can be significant even for small changes in this parameter (see below).
   The resulting profiles, assuming a  complete spherical atmosphere (the entire $\pm \pi/2$ radians range relative to the
   central plane), are shown in Fig.~\ref{fig:profiles_Ledd}. As expected, the profile becomes narrower with the increasing
$\Gamma$ reflecting the fact that, as the luminosity increases,  the cloud spend less and less time at small radii.

  The top part of Table~1 provides additional information about the 
  calculations. For each profile we give the FWHM in units of $v_{Kepler}(r_0)$, the mean
  emissivity weighted radius, $<r>/r_0$, and the mass correction factor 
   $f$ (eqn.~\ref{eq:old_mbh}). The calculation of $<r>$ is obtained by weighting the emissivity of the 
cloud and the time it spends at each radius.  This is roughly equivalent to
  the observed RM radius. The mass correction factor is obtained by requiring 
  $f FWHM^2 <r> /G = M_{BH}$. 
  We also show (in parenthesis) the values of FWHM and $f$ obtained for the case of a thick central disk
which represents only a part of a spherical distribution.
 Here the cloud distribution correspond to a width, relative to the
central plane, of $\pm \pi/4$ radians. The reduction in FWHM relative to the complete sphere is about a factor of
0.6 and there is a corresponding increase in $f$.
Computed line profiles that are typical of this and similar geometries are shown in Fig.~\ref{fig:profiles_sectors}.

\begin{table}
\begin{center}
\caption{Line widths, mass conversion factor $f$, and emissivity-weighted radii for various models\tablenotemark{a} }
\begin{tabular}{lccc}
$\Gamma$  & FWHM/$v_{Kepler}(r_0)$ & $<r>/r_0$ & $f$    \\
\hline 
$s=1.2$   & $r_{23}=10 r_0$       & $v_0=0.5$ & \\
\hline
0.05      &  1.58 (0.93)          &  0.54     & 0.75 (2.18)  \\
0.1       &   1.55 (0.92)         &  0.54     & 0.77 (2.21)  \\
0.3       &   1.45 (0.87)         &  0.56     & 0.85 (2.37)  \\
0.5       &   1.34 (0.81)         & 0.59      & 0.94 (2.56)  \\
0.7       & 1.15  (0.72)          & 0.68      & 1.11 (2.78)  \\
0.735     & 1.06   (0.68)         & 0.78      & 1.13 (2.76)   \\
\hline
$s=1.2$   & $r_{23}=10 r_0$ & $v_0=0.25$ & \\
\hline
0.05      &   1.04         &  0.45     & 2.05  \\
0.1       &   1.02         &  0.45     & 2.10  \\
0.3       &   0.95         &  0.47    & 2.39  \\
0.5       &   0.87         & 0.49     & 2.74  \\
0.7       &   0.76         & 0.52      & 3.31  \\
0.91      &   0.59         & 0.67     &  4.32   \\
\hline
$s=1.2$   & $r_{23}=r_0$ & $v_0=0.5$ &  \\
\hline
0.01     &  1.57                 & 0.54      & 0.76    \\
0.03     &  1.51                 & 0.55      & 0.80  \\
0.1      &  1.23                 & 0.64      & 1.03    \\
0.116    &  1.055                & 0.79      & 1.13    \\
\hline
\end{tabular}\end{center}
$^{\mathrm{a}}${Assuming the line
emissivity is strictly proportional to the cloud cross section and $\alpha(r)=0.5$.
In all cases $v_0=v_{\theta}(r_0)$. 
Numbers for $f$ assume spherical BLRs  
(numbers in brackets assume a $\pm \pi/4$ 
radians thick disk).}
\label{tab:profile_parameters}
\end{table}

\begin{figure}
\includegraphics[scale=0.42]{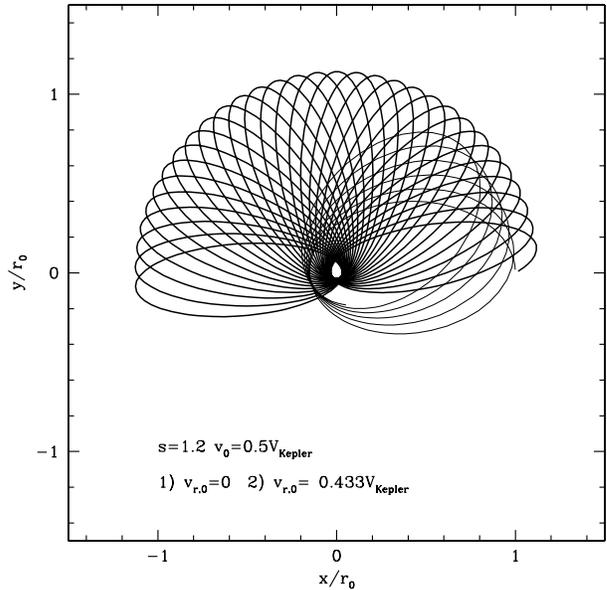}
\includegraphics[scale=0.42]{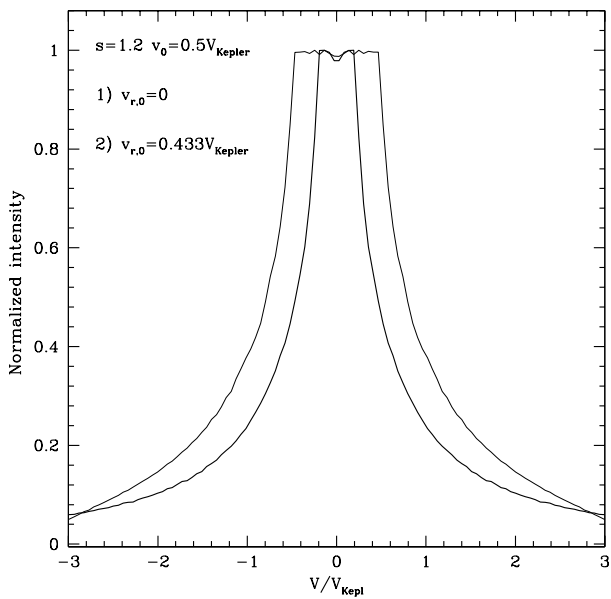}
\caption{Top: Planar orbits of two clouds with the same orbital energy and different angular momentum.  
The model parameters are $\Gamma=0.1$,
$s$ = 1.2 \mbh= 10$^{8}$ M$_{\odot}$ and other parameters  as in case 2) of Fig.~1.  The
less eccentric case ((1), thin line) corresponds to maximum angular momentum at $r = r_{0}$
 with  $v_{0} = 0.5$ and no initial radial velocity ($v_{r,0}=0$). 
The more eccentric case ((2), thick line) assumes $v_0=0.25$ and $v_{r,0}=0.433 v_{Kepler}$. 
Bottom: Line profiles for the two cases (same notation as in Fig.~2).
The narrower profile corresponds to  orbit 2).
}
\label{fig:different_orbits}
\end{figure}

\begin{figure}
\plotone{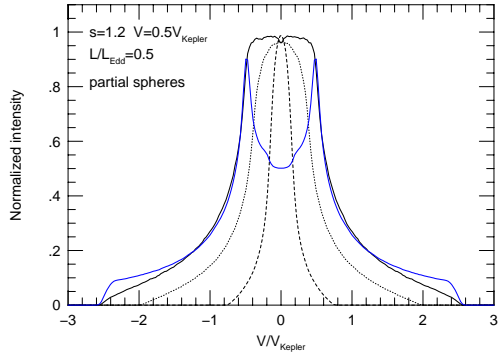}
\caption{Same initial conditions as in Fig.~\ref{fig:profiles_Ledd} for  $\Gamma=0.5$.
The various profiles represent motion in different spherically shaped atmospheres.
The narrowest profile (dashed line) represents a sphere where  clouds occupy only the section 
between -0.3 and +0.3 radians relative to the mid-plane (which is perpendicular to the line of sight). The other cases are for wider
coverage with clouds between -0.9 and +0.9 rad (dotted line) and -1.5 to +1.5 rad (solid line). The double
peak profile illustrates the case of two polar caps where the clouds occupy a sphere whose 
mid-section, between -1.2 and +1.2 rad, has been removed.
}
\label{fig:profiles_sectors}
\end{figure}

To explore models with different initial conditions, we computed two cases of planar orbits with the 
same orbital energy and different
angular momentum. One such example is shown in Fig.~\ref{fig:different_orbits}.
The less eccentric case in the diagram corresponds to the orbit labeled with 2) in Fig.~1 for which $v_0=0.5$.
The  more eccentric one assumes $v_{0}=0.25$ but with a non-zero radial velocity of $v_r=0.433$. This results in a much 
narrower profile. In the
middle part of Table~1, we report other cases where $v_{0}= 0.25$ and $v_r=0$. Such orbits are again very
eccentric and the profiles are, indeed, much narrower.
The corresponding values of $f$
are now larger by a factor of 2-3 than those observed. Additional models (not shown here) with
larger initial angular momentum, give larger FWHM and smaller $f$. Obviously, some combination
of all those is required to explain real observations.
The bottom part of Table~1 shows the results of a set of line profile
calculations carried out for  smaller
column density clouds. We chose $r_{23}=r_0$ which corresponds to
a factor of 6.3 decrease in  \Ncol\ relative to the case shown at the top of the table.
The scaling of FWHM between the two cases is simply by the corresponding
factor in $\Gamma$ (i.e. the same FWHM for $\Gamma$ smaller by a factor of 6.3). This illustrates the fact that in an atmosphere with a large range of column densities,
there are {\it always} clouds that are close to being ejected from the system at large distances.

The changes in  $<r>$ for a given $r_{23}$ shown in Table~1 are due to the fact that as $\Gamma$ increases,
and radiation pressure is more important, the clouds spend more and more time away from the BH.
This is noticeable for the case of $r_{23}=10 r_0$ when $\Gamma$ approaches 0.73
and  for the case of $r_{23}=r_0$ when $\Gamma$ approaches 0.1. 

As noted in \S1, RM campaigns show that \rblr(\hb)$ \propto L_{bol}^{0.6 \pm 0.1}$.
It is interesting to note that this behavior is not very different from what is calculated here
for  the changes in $<r>$ if we compare values over the range where $\Gamma$ approaches its limiting value. 
 However, it is {\it not} the case when $\Gamma$
changes by similar factors close to the lower range shown in the table, where radiation pressure is
negligible.

The values of $f$ computed here should be compared with those
determined observationally for selected AGN samples with measured $\sigma*$, in particular the Onken et al. (2004) and 
the Woo et al. (2010) AGN samples.
The simulations illustrate how this factor depends 
on the BLR geometry, the distribution of $\Gamma$ among objects in the sample and the distribution
of \Ncol\  in individual BLRs. 

An important point of the new calculation is the relatively little change in the value of FWHM listed in Table~1,
only a factor of $\sim 1.5$ over most of the range of $\Gamma$ except very close to the limiting value.
The changes in $f$ are also small, only a factor of $\sim 1.3$ over the same  range in $\Gamma$.
This seems to be in contradiction to the naive expectation that, for cases of increasing $L$,
the term  $<r> FWHM^2/G$ will deviate more and more from \mbh\ (e.g. eqn.~\ref{eq:M08_mbh}).
There are two reasons for this behavior. 
First, for realistic cases where \Ncol\ depends on the cloud location, the mean emissivity distance and the velocity depend
on $L/M$ rather than on $L$. This suggests that very low and very high luminosity AGNs with similar
$\Gamma$ will react to radiation pressure force in a similar way. 
Second, for a planar motion, the changes in the radial potential $\Phi(r)$ do not affect the cloud velocity in a linear way. In fact,
the mean orbital changes in $v_{\theta}$ are small enough such that the overall FWHM is very far from zero even for marginally
escaping clouds. Moreover, the mean cloud location, $<r>$, is increasing in reaction to the
increasing radiation pressure term. The end results is that the product 
 $f <r> FWHM^2/G$, with a constant value of $f$, 
is always a reasonable
approximation for \mbh\ with little dependence on the relative importance of gravity and radiation pressure force.
We return to this issue in \S4\ where we suggest a new way to evaluate \mbh\ taking into account radiation pressure
acceleration.

Finally, we note that while radiation pressure is negligible for very small
values of $\Gamma$, the $s$-dependence of the cloud properties is still very important.
For example, an
 $s=0$ atmosphere gives constant column density clouds (similar
to what was assumed in M08)
yet, the mean emissivity radius, the FWHM of the emission lines and
the mass correction factor $f$ in this case are always
different from those of the $s=1.2$ case, regardless of the column density.
The reason is
the dependence of the cloud cross section on $s$. For example, in the case of $\Gamma=0.01$
(first entry in the bottom part of Table~1), the $s=0$ case gives $<r>/r_0=0.39$ (compared with
0.53 for $s=1.2$) and FWHM$/v_{Kepler}=2.45$ (compared with 1.51). The resulting $f$ is therefore
much smaller (0.42 compared with 0.76).
Thus, the radial dependence of the cloud properties are important for all $\Gamma$.

\subsection{Applications to spectroscopic observations of AGNs}

The examples discussed above were normalized to give a typical \rblr(\hb) for AGNs
with \mbh=$10^8$\Msun\ and $\Gamma=0.1$. However, the computed line profiles 
 cannot be directly compared with the observations of such sources for several reasons.
 First, we only consider a situation involving one type of clouds and neglect 
the possibility of different populations under different
 physical conditions
 {\it in the same source}. This applies to the distributions of both  \Ncol\ and $U(r)$. 
For example, eqn.~\ref{eq:old_rblr} and the  constants given in \S1 suggest that, in general, 
$r_{BLR}(H_{\beta})$/$r_{BLR}(Civ \lambda 1549) \simeq 3$. The question is whether cloud distributions like
those considered in Table~1  can reproduce this ratio.
Second, we did not take into account changes in $\alpha(r)$, the fraction of \lbol\ which is absorbed by  
clouds at various distances. This can be an important factor close to the BH where clouds become 
partly transparent. In this case, much of the
Lyman continuum radiation is not absorbed and radiation pressure 
force is reduced.
It can also affect  
medium to large column density clouds at large distances where $\alpha(r)$ approaches unity.
For example, assuming $\alpha(r)=0.75$ instead
of $\alpha(r)=0.5$ in the calculations of Table~1 results
in a limiting value of $\Gamma$ which is about 0.4 
compared with $\Gamma=0.735$ listed in the table.

\begin{figure}
 \plotone{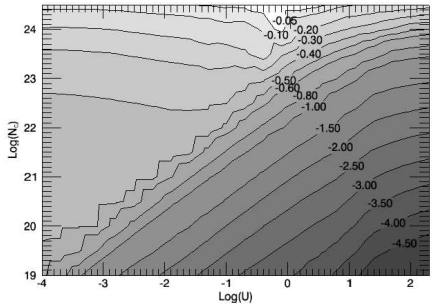}
 \caption{Part of the $\alpha(r)$ grid (fraction of the total continuum flux absorbed by the clouds)
used in the present calculations. Numbers along the contour lines are $log\,\, \alpha$.
 }
 \label{fig:grid_alpha}
\end{figure}

To illustrate these effects, and to provide more realistic line profiles, 
we computed two grids of photoionization models for a range of column density
and ionization parameter using the code ION (Netzer 2006).
The first grid supplies calculated line intensities for  \hb\ and \civ\  over a large range in $U(r)$.
Given $r$ from the cloud motion simulation, we use the grid to compute $j(r)$ and thus a more realistic
 $\epsilon(r)$.
The second grid supplies the absorbed fraction, $\alpha(r)$,
as a function of $U(r)$ and \Ncol. 
Fig.~\ref{fig:grid_alpha} shows part of the $\alpha(r)$ grid to illustrate the expected range in this parameter.
We have not included the changes in gas density  since they do not
 play a major role over the range of conditions considered here. We have also not considered 
anisotropy of the line emission which
is bound to have an effect on the FWHM of some lines. Such modifications will
be included in a  forthcoming paper that is intended to present a comparison with observed line profiles.

We tested a large number of single-zone models using the above grids of 
$U(r)$ and $\alpha(r)$. The models cover a large range in angular momentum 
and BLR geometries. We have specifically investigated three cases of different eccentricity, defined by
three values of $v_0(r_0)$, 0.25, 0.5 and 0.75. These were calculated with different $\Gamma$ and $r_{23}$.
In general, it is easy to reproduce the observed I(\civ)/I(\hb) but difficult to account,
at the same time, for the emissivity weighted radii of the two lines (eqn.~\ref{eq:old_rblr})
and  the line width ratio. For example, the best case of the three, with $v_0(r_0)=0.25$, gives I(\civ)/I(\hb)=4.4,
$<r>$(\civ)/$<r>$(\hb)=0.67 and FWHM(\civ)/FWHM(\hb)=1.43.
The conclusion is that, within the range of parameters assumed here, there is no obvious way to explain all those properties
when keeping with the idea of a single column density distribution (i.e. a single $r_{23}$).

\begin{figure}
 \plotone{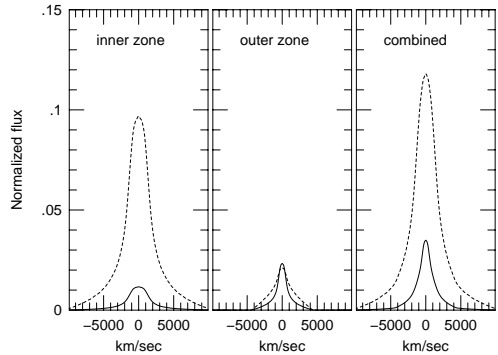}
\caption{Calculated \hb\ (solid line) and \civ\ (dashed line) profiles for a two zone model. 
Left: line profile for the inner zone.
Middle: line profiles for the outer zone. 
Right: The combined line profile. For FWHMs and general normalization see Table 2.
}
\label{fig:two_zone_profiles}
\end{figure}

We also tested a case of \mbh$=10^8$\msun, $\Gamma=0.1$ and {\it two 
distinct} cloud populations in  inner and  outer zones with some overlap between the two. 
In this case, the initial conditions for the two populations are decoupled from each other
but the changes in density, column density and ionization parameter follow the same pattern
with the same $s=1.2$ density law.
The FWHM of both emission lines were calculated under the assumption of
a thick central spherical sector with clouds occupying a region of $\pm \pi/4$ radians relative to the
central plane. 
The inner zone clouds have  $r_0=5 \times 10^{16}$ cm and $U(r_0)=10^{-1}$ and
the outer-zone clouds 
$r_0=3 \times 10^{17}$ cm and $U(r_0)=10^{-2.5}$. The starting velocity in both zones is
$v_0=0.5$ at the appropriate  $r_0$. In both zones $r_{23}=3 r_0$.
We followed the cloud motion and calculated, in each zone, the
line intensity ratio, I(\civ)/I(\hb), the line FWHMs, and the emissivity weighted radii. These
numbers are listed in Table~2 where we also show the properties of the combined spectrum which is
calculated under the assumption of equal contributions to \hb\ from both zones. The emissivity weighted radii
for the two zones are given
in units of the RM-radii of the two lines (eqn.~\ref{eq:old_rblr}).
 This very simple two-zone model gives results that are in good agreement
with the observations of many low-to-intermediate luminosity AGNs.
Fig.~\ref{fig:two_zone_profiles} is a graphical summary of these results. The left and central  panels show
\hb\ and \civ\ profiles for the inner and outer zones, again assuming isotropic line emission, and the right panel shows the 
combined two-zone profiles.
\setlength{\tabcolsep}{.5pt}
\begin{table}
\begin{center}
\caption{Properties of the two-zone model with $v_0(r_0)=0.5$.}
\begin{tabular}{lccccc}\hline
 Zone & FWHM(\hb)      & FWHM($Civ$)     &  $\frac {r(\hb)}{r(RM,\hb)}$ & $\frac{r(Civ)}{r(RM,Civ)}$ & $\frac{I(Civ)}{I(\hb)}$  \\
      &(km\,s$^{-1}$) & (km\,s$^{-1}$) &     &     &         \\
 \hline
Inner       &  3160    & 3450    & 0.32     & 0.88   & 9.55 \\
  Outer     &  1390   &   2580  &  1.64   & 3.2  & 1.45   \\
Combined   &  2060  &   3390  &   0.98   & 1.1  & 5.5   \\
\hline
\end{tabular}
\end{center}
\label{tab:hb_civ}
\end{table}

In conclusion,
the simple single zone models explored here cannot reproduce all
the observed properties: line intensity ratio, mean emissivity radii and  
FWHM ratio. The main reason is that the starting conditions fix the cloud orbit, and hence
the line emissivity and FWHM.
Simple two-zone models like the ones presented here can account for most observed properties of the \hb\ and \civ\ lines.
In particular, they can
account for the mean line ratio, the mean emissivity weighted radii and the mean relative
FWHM of the \hb\ and \civ\ lines measured in various RM samples.
Obviously, such simple models do not intend to explain all the observed line profile properties
that can differ from one object to the next and contain additional components
(see some obvious examples
for complex \civ\ profiles in Richards et al. 2002 and Sulentic et al. 2007).
Fitting those is deferred to a forthcoming paper.
 
\section{Discussion}
\subsection{General considerations}
The above calculations allow us to investigate the intensity, the
width and the
shape of the broad emission lines and to evaluate various methods used
to estimate \mbh. 
We defer the  discussion of specific observed line profiles to a future paper.

Assume a system of clouds with a given total amount of gas and a large range of column densities. Such
a system will eventually break into three: virialized clouds, non-virialized bound clouds
and escaping clouds.
The third group will not contribute significantly to the observed line emission
for more than  several dynamical times. 
The relative contribution of the first and second groups to the line emission depend
on the cloud mass distribution. 
A sudden increase in \Lbol\ will increase the importance of radiation pressure and will remove more
gas from the system. A decrease in \Lbol\ will drive the system closer to  virial equilibrium.
 A new gas supply, e.g. from a disk-wind, will produce  bound as well as unbound clouds.
All aspects of this general scenario must be considered when evaluating the observed line profiles and
the various methods developed to use them in estimating \mbh.

A major objective of the present paper is to evaluate the accuracy and the normalization of various
\mbh\ estimators in
type-I AGNs.  The results presented in Tables 1 \& 2  suggest the following:
\begin{enumerate}
\item
Every AGN is likely to contain a large number of clouds with a large range in \Ncol. This can be  the result of a broad
cloud mass distribution and/or due to 
  cloud motion in a radial-pressure dependent environment with a positive value of $s$.
A given $\Gamma$ results
in a lower limit on \Ncol\ at {\it a given location} for  a given orbit eccentricity.
Under such conditions, there are always some 
clouds, e.g. those that are very close
to the BH, for which radiation pressure is negligible. For others,
radiation pressure can be very important.
\item
For a small enough \Ncol, the effective \rblr\ depends on both $\Gamma$ and \Ncol. Under
these conditions, the BH mass itself is an important factor in determining \rblr.
To illustrate this, consider two AGNs with identical SED, \Lbol, BLR geometry, \Ncol\ distribution and inclination to the
line of sight. The effective \rblr\ in the two
is the same provided they harbor identical BHs. Different \rblr\ will be measured if the two BHs have different
masses despite of the fact that \Lbol\ is the same in both. This is the result of the larger $\Gamma$
in the smaller BH AGN. The effect may not be recognized in a large sample of sources and can, in fact, be attributed to 
a large intrinsic scatter in the $L_{bol}-r_{BLR}$ relationship. 
Any derived 
 $L_{bol}-r_{BLR}$ relationship  will depend on the properties of the sources in the chosen RM sample, 
 in particular on the distribution of $\Gamma$.
\item
Assuming a range in \Ncol\ in {\it every} AGN, the M08 suggestion to include a luminosity dependent term in the calculation of
\mbh\ (eqn.~\ref{eq:M08_mbh}) is not in accord with our  calculation that indicate that
\rblr\ and FWHM  depend on $L/M$ and not on $L$. 
\end{enumerate}

The multi-year RM campaign of NGC\,5548 is the best example to test some of these ideas in a specific source.
The campaign has been described and analyzed in numerous papers and the ones most relevant to the present
study are Peterson et al. (1999) and Gilbert and Peterson (2003).

\begin{figure}
 \plotone{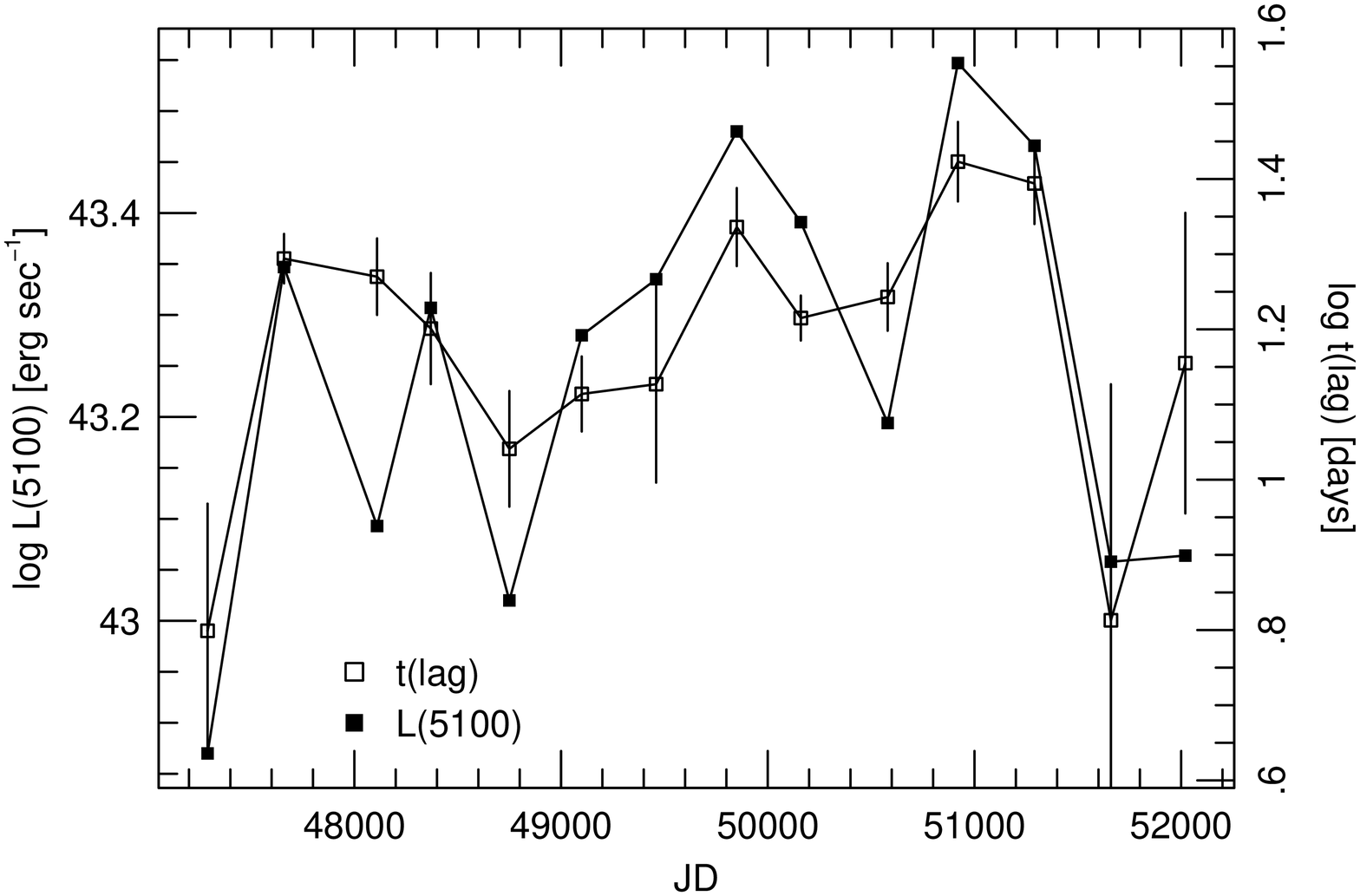}
\caption{Changes in continuum luminosity (\Lop) and time lag for NGC\,5548
(data from Bentz et al. 2009). Error bars on \Lop\ were omitted for clarity.
}
\label{fig:n5548_l_t_jd}
\end{figure}

\begin{figure}
 \plotone{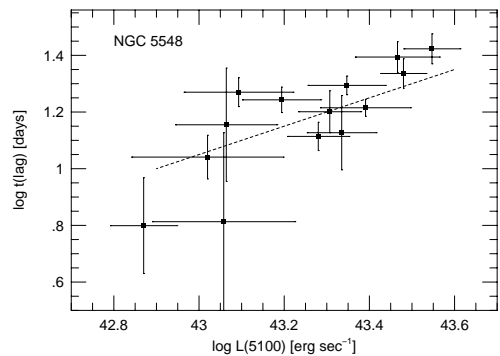}
\caption{The correlation of \Lop\ vs. t(lag) for NGC\,5548. Data as in fig.~\ref{fig:n5548_l_t_jd}.
The dashed line has a slope of 0.5.
}
\label{fig:n5548_l_t}
\end{figure}

Fig.~\ref{fig:n5548_l_t_jd} shows the variations in \Lop\ and time lag (in this case the centroid of the CCF) in NGC\,5548.
Each point represents a full observing season which is typically $\sim 300$ days long. 
The data are taken from the recent compilation of Bentz et al. (2009) which provides the best galaxy subtracted
flux at 5100\AA. The uncertainty on \Lop\ is basically the range of this quantity over the observing season. This
is of the same order as the variation from one season to the next. As clearly seen from the diagram, \rblr(\hb)
lags the continuum in such a way that more luminous phases are associated with longer lags. This has been noted in earlier
publications, e.g.  Gilbert and Peterson (2003). Fig.~\ref{fig:n5548_l_t} shows t(lag) vs. \Lop\ for the same data set. 
While the uncertainties are large, some correlation, with a slope of 0.5-1, is evident. 
An earlier version of the diagram, with fewer points, is shown in Peterson et al. (1999). 

For NGC\,5548, \mbh$\simeq 10^8$ \msun\ and \rblr(\hb)$\simeq 20$ l.d.. Thus, the dynamical time is
of order 6 years and the time it takes to change \rblr\ by 50\% (e.g. Table 1) is approximately
3 years. This seems to be compatible with the changes in \Lop\ and 
t(lag) in fig.~\ref{fig:n5548_l_t_jd}, thus some adjustment of \rblr(\hb) due to the effects discussed in this work
are possible.
The measured \Lop, with a bolometric correction factor of about 10, indicates a mean $\Gamma$  of
about 0.02. The bottom part of Table~1  provides approximate  parameters for such a case.
Any successful model of NGC\,5548
should account for the behavior shown in Fig.~\ref{fig:n5548_l_t_jd}, as well as for the observed FWHMs  and luminosities of both H$\beta$\ 
and \civ. While the full investigation is deferred to a future paper, we consider here the predicted lags 
for $\Gamma$ =0.005, 0.01, 0.02 assuming a BH mass of 10$^{8}$ M$_{\odot}$ and
 two families of clouds: one with $r_{23} = 10 r_{0}$ and $v_{0}=0.5$, and one with $r_{23} = 0.093 r_{0}$ and $v_{0}=0.25$. 
The second assumed family of clouds results in pseudo-orbits of higher eccentricity that, as explained earlier,
are more strongly affected by radiation pressure. 
In both cases, the predicted lags for $\Gamma  = 0.02$  are consistent with the observed values 
($\log t $(lag) $\approx$ 1.34 at $\log L_{5100} \approx 43.48$).
However, the calculated slope of $\log t $(lag)  vs. \Lop\ is flatter than observed.

As argued earlier, a single family of BLR clouds cannot provide a full explanation to the observed
spectrum of many AGNs. This must applied to NGC\,5548 (to appreciate the complexity of this case see the various
components considered by Kaspi and Netzer (1999) to explain only the variable line intensities).
The simple examples considered here suggests that dynamical scaling
of the BLR in NGC\,5548, due to radiation pressure force, is an additional, physically-motivated mechanism that must be
added to any cloud model when attempting to explain the observed 
variations in $t$(lag).

\subsection{Evaluation of present \mbh\ estimators}

Current BH mass estimates utilize RM-based measurements of \rblr, measured FWHMs (or an equivalent velocity estimator) 
of certain broad emission lines, and eqn.~\ref{eq:old_mbh}.
The normalization constant $f$ is obtained by a comparing \mbh\ obtained in this way  with the mass obtained from the \msig\ method.
Having examined a large range of cloud orbits and line profiles under various conditions, and the corresponding values of the effective \rblr,
we are now in a position to evaluate the merits of this method.

We consider three general possibilities.
 The first is the case where all AGNs contain BLR clouds with a wide column density distribution.
A  randomly chosen object will
have in its BLR some clouds that are affected by radiation pressure force and others that are not. This is
the case for {\it any} $\Gamma$. The  cloud dynamics and the observed line profiles reflect the (unknown) 
column density distribution. Our calculations suggest that an RM sample drawn randomly from such an AGN population can be 
safely used to determine the best value of $f$  by comparing the derived \mbh\  with 
the \msig\ method.
This is justified by the fact that $M_{BH} \propto <r>FWHM^2$ even if
radiation pressure force is important (see Table~1).
 The observed FWHMs are, indeed, smaller than the ones that 
 would have been observed if all clouds had extremely large column densities. This, however, has no
practical implication since the column densities are not known and $f$ is simply a normalization factor
that serves to bring two completely different methods of estimating \mbh\ into agreement.
Mass estimates obtained in this ways are reliable provided the properties of the RM sample represent well the population properties.

The second case reflects a situation where the cloud column density distribution is, again, very broad but
part of the population is under-represented in the
RM sample. For example, the RM sample may contain mostly sources with $\Gamma \sim 0.1$ while the overall distribution of $\Gamma$
is much wider.
 In this case, the normalization factor $f$
will reflect only the properties of the measured sources and its use will provide poor  mass estimates for 
cases with much larger or much smaller accretion rates. This may well be the case in the RM sample  which is most
commonly used (Kaspi et al. 2000; Bentz et al, 2009) that
contains only very few AGNs with $\Gamma > 0.3$. The numbers in Table~1
 enable us to evaluate the resulting deviations in the estimated \mbh. For example, if we use the first part of the table 
and assume a source with a certain \Lbol\ and $\Gamma=0.1$,
we find that the mass of a similar \Lbol\ source with $\Gamma=0.7$ will be under-estimated
by a factor of 1.11/0.75.

Regarding the second case, it is important to note that under-estimates and over-estimates of \mbh\ are
equally likely. Consider again an  RM sample where, for most sources, $\Gamma=0.1$. This results in
a certain value of $f$ which takes into account
the effect of radiation pressure force in {\it some} of these sources
(see bottom part of Table~1).
Assume a second, randomly selected AGN sample with a similar BH mass distribution
but a typical $\Gamma$ which is much smaller than 0.1. 
Most measured FWHMs in this sample are {\it broader} than those
in the RM sample because radiation pressure force is not as effective in reducing the cloud velocity.
Using the value of $f$ derived for the RM sample will result in {\it over estimating}
\mbh\ in the second sample. The lower part of Table~1 gives some idea about the magnitude of this effect, e.g. an over estimate by a
factor of 1.01/0.76.

The third case is similar to the first one except that large luminosity variations, on time scales
that are not too different from the BLR dynamical time, are occurring in most sources, including
those selected for RM monitoring. Table~1 shows that, like the first case, the deduced $f$ represents
well the population because $<r>$ follows the variations in \Lbol.
The  mean \mbh\ in such a sample is recovered albeit with a larger uncertainty.

\subsection{Alternative \mbh\ estimators}

Given the above considerations, we now investigate an alternative method to calculate
\mbh. The method takes into account the effect of radiation pressure force on the cloud motion
and the results will be compared to those obtained by the old method (eqn.~\ref{eq:old_mbh})
and by the M08 method.

Our new calculations indicate that the emissivity weighted \rblr\
depends both on the (large) range in $L$ across the entire AGN population, as well as on short time scale changes in \rblr\
in individual  sources. The first of those depends roughly on $L^{1/2}$ and is
a manifestation of the observational fact that the ionization parameter, $U(r)$, and the spectral energy distribution (SED),
 are not changing much with source luminosity. 
The second reflects changes in the BLR structure due to the  reaction of various column density clouds to the (changing)
radiation pressure force. This  depends on
both \Lbol\ and \mbh. This is seen for example in eqn.~\ref{eq:r_min} for the critical radius where clouds can escape
the system and also in the calculations of Table~1.
It is therefore reasonable to assume that \rblr\ is given by an expression of the form,
\begin{equation}
r_{BLR}= a_1 L^{\gamma} + a_2 (L/M)^{\delta} \,\, ,
\label{eq:new_rblr}
\end{equation}
where $a_1$ and $a_2$ are constants and $L$ is a measure of the source luminosity, e.g. \Lop\
if \rblr=\rblr(\hb).
Obviously, the above approximation is not unique and one can assume other dependences that are consistent
with the line profile calculations, e.g. a dependence of FWHM on $L/M$.

The idea of introducing a second, luminosity dependent term into the  calculation of \mbh\ is not new.
In particular, M08 suggested
 an expression for \mbh\ which depends on both $L^{1/2}$ and $L/N_{col}$ (eqn.~\ref{eq:M08_mbh}).
Assuming all AGNs obey  the same relationship, and \Ncol\ is the same in all, 
the M08 expression leads to extremely large values of \mbh\ for the most luminous AGNs.
The reason is the linear dependence of \mbh\ on $L$ at very high luminosities combined with the calibration of the
 relationship at small $L$, typical of the \msig\ sample of Onken et al. (2004).
The additional consequence of this approach is an upper limit of $\Gamma \sim 0.1$ in many
high luminosity, large BH mass sources. 
In their later work, M09 considered the possibility that \Ncol\ can differ from one
source to another but is still constant for all clouds in a given BLR. This would result in smaller \mbh\ and
larger $\Gamma$ in some high luminosity sources since in 
 some BLRs, \Ncol\ can exceed $10^{23}$ \cmii\ by a large factor
thus reducing the importance of radiation pressure force.

The limitation of the M08 mass estimate is the detachment of $L$ from \mbh. As shown
here, this is not the case in more realistic BLRs, especially those where the masses of the clouds are conserved.
In such cases, the location of the outer 
clouds  that still contribute to the line
profiles  depends on $L/M$ and the 3D-velocities of the marginally bound clouds are such that
the product \rblr$ FWHM^2$ is not very different from what is found in pure gravity dominated systems.
Moreover, for pressure confined clouds, the dependence on \Ncol\ is likely to be different in different parts of the BLR.
Thus, we are looking for an expression that will reflect, properly, all these effects and
will allow for the possibility of a range of column densities in {\it every source}. 
We also want to avoid biasing in the derivation of \mbh\ 
in the limits of very large or very small $L$ and to retain the experimental results 
that $r_{BLR} \propto L^{\gamma}$
with $\gamma = 0.6 \pm 0.1$.

All the above can be achieved by assuming that \rblr\ is given by   
eqn.~\ref{eq:new_rblr} and requiring that $M_{BH} \propto r_{BLR} FWHM^{2}$. 
For the sake of simplicity, we assume $\gamma=0.5$ and $\delta=1$ and
substitute  eqn.~\ref{eq:new_rblr} into the mass expression.
This leads to a simple quadratic equation
in \mbh\ with the following  solution,
\begin{equation}
M_{BH}=\frac{1}{2} a_1 L^{1/2} FWHM^2
       \left[ 1 + \sqrt (1+ \frac{4 a_2}{a_1^2 FWHM^2}) \right ] \,\, ,
\label{eq:new_mbh}
\end{equation}
where $a_1$ and $a_2$ are the same ones used in  
eqn.~\ref{eq:new_rblr} except for a common multiplicative constant
which depends on the units of \rblr, \L5100\ and \mbh. For example, using the measure parameters for the \hb\ line, 
$L=$\Lop, FWHM=FWHM(\hb), then  the constant multiplying $a_1$ and $a_2$ in 
eqn.~\ref{eq:new_rblr} is 10$^{16.123}$ when  
\mbh\ is measured in \msun,  \L5100\ in units of $10^{44}$ \ergs\ and \rblr\ in cm.

We used eqn.~\ref{eq:new_mbh} and the Woo et al. (2010) sample to find $a_1$ and $a_2$ for 
29 AGNs with measured $\sigma*$. The list is an extension of the one used by Onken et al. (2004) that
contains only 16 sources. We have supplemented the data in Woo et al. by data from Bentz et al. (2009) on \rblr\ and \L5100\
where this information was missing.
First, we performed a $\chi^2$ analysis on \mbh(RM) vs. \msig\ using the parameters recommended by G{\"u}ltekin et al. (2009).
This gave
$f=1.0$ which is consistent with the values found by Onken et al. (2004) and Woo et al. (2010)\footnote{Onken et al.
(2004) and Woo et al. (2010) carried the analysis using the \hb\ line dispersion rather than FWHM(\hb).
For the sample in question, this line-width measure is smaller than the FWHM(\hb) by a factor of approximately 1.9 
 leading to a corresponding increase in the mean $f$ by a factor of about $1.9^2$.
All these numbers are sensitive to the error estimate in $\sigma*$ and in the virial product.}

Next we carried out a $\chi^2$ minimization to solve for $a_1$ and $a_2$ in eqn.~\ref{eq:new_mbh}.
 Since the minimization involves the error estimate
on \L5100, and since this error is not very well defined given the combination of observational uncertainly
and the intrinsic scatter in \L5100\ over several long
RM campaigns, we decided to adopt
a uniform value of $\Delta L_{5100}/L_{5100}=0.3$. We also assume a minimum of 0.1 to 
$\Delta(\sigma*)/\sigma*$ and a minimum of 0.05 on $\Delta(FWHM)/FWHM$. Our results depend slightly on these assumptions.

The best values obtained in this procedure are 
$a_1=4.1$, $a_2=7.1 \times 10^7$ and $\chi^2/{\nu} = 1.73$. Extensive tests show that the $\chi^2$ changes very little if
$a_1$ or $a_2$ are changing by up to 10\%. This is the result of some degeneracy between  $a_1$ and $a_2$
  (see eqn.~\ref{eq:new_mbh}). The average deviation between the new mass estimates and those obtained by the $M-\sigma*$ 
method is 0.31 dex. There is a weak dependence of the deviation on the line width (larger deviation for larger FWHM(\hb))
which is marginal given 
the small number of sources in the sample. The corresponding number for the deviation of masses obtained directly from the RM measurements and the
above value of $f$ is 0.36 dex. Thus the new method is, indeed, superior in this respect.
 Obviously it is not surprising to find such an improvement when adding a new free parameter to
the model.

To compare the various mass estimates more thoroughly, 
we calculated \mbh\ in three different ways: the old method (eqn.~\ref{eq:old_mbh}) with $f=1.0$,
the M08 method (eqn.~\ref{eq:M08_mbh}) with $f_1=0.56$ and $f_g=10^{7.7}$ (as in M08), 
and the new method (eqn.~\ref{eq:new_mbh}) with
the above $a_1$ and $a_2$. For the M08 method, we followed the M09 recommendation and
assumed a log normal distribution of \Ncol\ with 
a mean of $10^{23}$ \cmii\ and a large standard deviation of 0.5 dex.
We also calculated \rblr\ in the old (eqn.~\ref{eq:old_rblr}) and new (eqn.~\ref{eq:new_rblr}) ways.

\begin{figure}
 \plotone{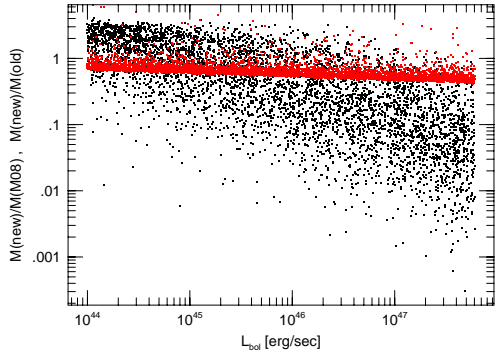}
\caption{Comparison of the various methods for calculating \mbh. The ratio of the new-to-old (red) 
and new-to-M08 (black) methods
are shown as a function of \Lbol\ for the simulated sample described in the text. Note the good agreement
between the old and the new methods and the large deviation from the method described in M08 for very large
and very small values of \Lbol.
}
\label{fig:simulation}
\end{figure}

Fig.~\ref{fig:simulation} compares two mass ratios, \mbh(new)/\mbh(old) (red points) 
and \mbh(new)/\mbh(M08) (black points),
in a large simulated AGN sample. The sample covers, uniformly, the luminosity range \Lop=$10^{43}-10^{47}$ \ergs\
and the simulations assume a Gaussian, luminosity independent distribution of FWHM(\hb) with a mean of 
4,500 \kms\ and a variance of 1,500 \kms.
The diagram shows that the new and old estimates are similar at all \Lbol\ but \mbh(M08) deviates from both, by
a large factor, at both low and high luminosities.
Moreover, the slight deviation between the new and old methods at the very high luminosity end, 
by up to about 0.2 dex in \mbh, is most likely due to the fact that the
procedure used to obtain  $a_1$ and $a_2$ is based on a sample of 29 mostly low-to-intermediate luminosity
AGNs while the simulations reach a much larger value of \Lbol. A comparison of the estimated \rblr\ (eq.~\ref{eq:old_rblr} and \ref{eq:new_rblr}) leads to
similar conclusions. 

We also made a similar test on the Netzer and Trakhtenbrot (2007) sample using all three methods. The luminosity
range in this case is smaller but the FWHM(\hb) distribution more typical of observed AGNs. The results (not
shown here) are very similar to those of the simulated sample.

In conclusion, the new method for estimating \mbh\ gives results that do not deviate much from the
old method which is based on a single constant $f$. This is true at both high and low luminosities and
over a large range in FWHM. Obviously, the range of parameters tested here ($s$, orbit eccentricity,
several types of cloud distributions, etc.) is rather limited and more extensive modeling is required
to confirm these results. However, it is our opinion that the main limitation of the \mbh\ determination
methods remains observational and is related to the fact that the present AGN \msig\ sample is small (29 sources) and cannot  
possibly represent the entire range of properties, mostly $\Gamma$,
observed in AGNs.

\section{Conclusions}

We have investigated the motion of BLR clouds with time-independent mass under a range of conditions
defined by a radial-dependent confining pressure.
These conditions enforce a range of \Ncol\ in every BLR, even if the intrinsic mass distribution of the cloud is narrow.
 We calculated cloud orbits under a central potential that includes a radiation
pressure term. The orbits were then combined to predict emission line profiles in several simple situations. We only considered
uniformly emitted emission lines and the preliminary comparison with with actual observations used realistic emissivity and
column density distributions but was limited to the \hb\ and
\civ\ lines and at most two different cloud distributions.
 We found  significant changes in cloud locations and velocities for those cases where the column densities are small enough to
allow a significant contribution due to radiation pressure.  This can be important in both high and low $\Gamma$ sources. 
However, while cloud orbits are strongly influenced by the radiation pressure force, there is a relatively small
change in the mean \rblr$ FWHM^2$ and hence no large underestimation or overestimation of \mbh. We illustrate this behavior in several cases but 
note that other cloud distributions, with different mass, location and velocity distributions, may lead to
somewhat different conclusions. 
We used the new results to suggest a novel method for calculating \rblr\ and \mbh\ by applying two new constants that were calculated by a comparison
of the \hb\ and \L5100\ observations and the $M-\sigma*$ AGN sample of Woo et al. (2010). We applied the method to several
large observed and simulated AGN samples and demonstrated good agreement between the new and the old, pure gravity based methods.
The comparison with the M08 methods shows large 
deviations in the estimates of \mbh.

\acknowledgements
We acknowledge useful comments by an anonymous referee and a detection of a typo in  Table~1  by J.M. Wang.
Funding for this work has been provided by the Israel Science Foundation grant 364/07 and by the Jack Adler Chair for Extragalactic Astronomy.
HN thanks the hospitality of Imperial College London and University College London where part of this work has been done. 
PM is grateful for the hospitality and support of the school of Physics and Astronomy at Tel Aviv University.

\end{document}